\documentclass[times, final, 3p]{article}
\newcommand{\mytitle}{Mean field homogenization schemes for composites with prolate and oblate spheroids: use of the orientation tensors and computation of the strain second-moments}

\title{\mytitle}

\usepackage{parskip}
\usepackage[fleqn]{amsmath}
\usepackage[top=3cm,bottom=4cm]{geometry}

\usepackage{amssymb}
\usepackage{amsthm}

\usepackage{bm}
\usepackage{tikz}

\usepackage{xcolor}

\newcommand{\dbldot}{\mathbin{\mathord{:}}}

\newcommand{\tens}[1]{\vec{#1}}
\renewcommand{\vec}[1]{\bm{\mathrm{#1}}}
\newcommand{\Eb}{\overline{\tens{\varepsilon}}}

\newcommand{\beq}{\begin{equation}}
\newcommand{\eeq}{\end{equation}}
\newcommand{\beqal}{\begin{equation}\begin{aligned}}
\newcommand{\eeqal}{\end{aligned}\end{equation}}
\newcommand{\tsigma}{\tens{\sigma}}
\newcommand{\tepsilon}{\tens{\varepsilon}}

\newcommand{\deriv}[2]{\dfrac{\partial #1}{\partial #2}}

\begin{document}

\maketitle
\begin{center}
\author{A. Martin$^a$}

\medskip

\author{\small{$^a$\textit{CEA, DES, IRESNE, DEC, SESC, LMCP, Cadarache, F 13108 St Paul Lez Durance, France}}}
\end{center}

\section{Introduction}
In this document we provide the homogenized stiffnesses from Mori-Tanaka scheme and Ponte-Casta\~neda and Willis scheme applied to
composites with spheroidal inclusions. The inclusions can be prolate spheroids ('fibers') or oblate
spheroids ('penny-shapes'). We show how to compute the homogenized stiffnesses for particular distribution
of spheroid orientations, using the second-order and fourth-order orientation tensors. We also provide formulas
to compute the derivatives of these quantities w.r.t. the material parameters, which is of particular interest for computing
the second-moments of the strains.

We consider a two-phase composite with isotropic phases, defined by the moduli $k_0,\mu_0,k_1$ and $\mu_1$, and we note $\tens C_i=3k_i\tens J+2\mu_i\tens K$ ($i=0,1$). The phase 0 corresponds to the matrix phase, whereas phase 1 corresponds to the inclusions. The volume fraction of inclusions is $f$.

The average of a quantity over all the inclusions is noted with $\langle.\rangle$. The average of a quantity over the entire material is noted $\langle.\rangle_{\Omega}$.
The average on the matrix is noted $\langle.\rangle_0$, and we also note $\langle.\rangle_1$ the average over the inclusion phase.

\section{Mori-Tanaka scheme and PCW scheme}
\subsection{Mori-Tanaka scheme}
In Mori-Tanaka scheme \cite{Mori}, the field in a spheroidal inclusion defined by axis $\vec n$ is given by
\beqal
\tepsilon_1(\vec n)=\tens A_1(\vec n)\dbldot\tens E_0,\qquad\text{with}\quad\tens A_1(\vec n)=\left[\tens I+\tens P(\vec n)\dbldot\left(\tens C_1-\tens C_0\right)\right]^{-1}
\eeqal
where $\tens P_0(\vec n)$ is the Hill tensor associated to the inclusion and the elasticity $\tens C_0$ and $\tens E_0$ is defined by the following equation
\beqal
\Eb=\tens E_0+f\,\left(\langle\tens A_1\rangle-\tens I\right)\dbldot\tens E_0\quad\text{so that}\quad\tens E_0=\left[\tens I+f\,\left(\langle\tens A_1\rangle-\tens I\right)\right]^{-1}\dbldot\Eb
\eeqal
The strain field in the matrix is considered equal to $\tens E_0$.
The average of the stress on the entire volume is then given by
\beqal
\langle\tsigma\rangle_{\Omega}&=\langle\tens C\dbldot\tepsilon\rangle_{\Omega}=(1-f)\,\tens C_0\dbldot\tens E_0 + f\,\tens C_1\dbldot\langle\tens A_1\rangle\dbldot\tens E_0\\
&=\left[(1-f)\,\tens C_0 + f\,\tens C_1\dbldot\langle\tens A_1\rangle\right]\dbldot\left[\tens I+f\,\left(\langle\tens A_1\rangle-\tens I\right)\right]^{-1}\dbldot\Eb
\eeqal
so that the homogenized stiffness given by Mori-Tanaka scheme is 
\beqal
\label{MT_stiffness}
\tens C^{\mathrm{MT}}&=\left[\tens C_0 + f\,\left(\tens C_1\dbldot\langle\tens A_1\rangle-\tens C_0\right)\right]\dbldot\left[\tens I+f\,\left(\langle\tens A_1\rangle-\tens I\right)\right]^{-1}
\eeqal
We hence see that the average $\langle\tens A_1\rangle$ is necessary to compute this homogenized stiffness.

\subsection{PCW scheme}
We assume that the distribution of the centers of the inclusions is given by a unique tensor $\tens P_d$.
This tensor is the Hill tensor relative to the distribution of inclusions (see \cite{PCW95}).
Hence, if this tensor is spheroidal, it is defined by an aspect ratio and a direction $\vec n_d$.
This direction hence defines a Walpole basis related to the \emph{distribution} of spheroids.
This basis must be distinguished from the basis defined by each normal vector $\vec n$ associated with a spheroidal inclusion.

\medskip

Now, we have for an inclusion whose axis is defined by $\vec n$:
\beqal
\tepsilon = \Eb - \tens P(\vec n)\dbldot\tens \tau + f \tens P_d\dbldot\overline{\tens\tau}
\eeqal
where $\tens \tau = \left(\tens C_1-\tens C_0\right)\dbldot\tepsilon$ and $\overline{\tens\tau}=\langle\tens\tau\rangle$. Hence,
\beqal
\left(\tens I+\tens P(\vec n)\dbldot\left(\tens C_1-\tens C_0\right)\right)\dbldot\tepsilon = \Eb + f \tens P_d\dbldot\overline{\tens\tau}
\eeqal
and
\beqal
\tepsilon=\tens A_1(\vec n)\dbldot\left[\Eb + f \tens P_d\dbldot\overline{\tens\tau}\right]
\eeqal
and
\beqal
\overline{\tens\tau}=\left(\tens C_1-\tens C_0\right)\dbldot\langle\tens A_1(\vec n)\rangle\dbldot\left[\Eb + f \tens P_d\dbldot\overline{\tens\tau}\right]
\eeqal
so that
\beqal
\left[\tens I-f\,\left(\tens C_1-\tens C_0\right)\dbldot\langle\tens A_1(\vec n)\rangle\dbldot\tens P_d\right]\dbldot\overline{\tens\tau}=\left(\tens C_1-\tens C_0\right)\dbldot\langle\tens A_1(\vec n)\rangle\dbldot\Eb
\eeqal
and
\beqal
\overline{\tens\tau}=\left[\tens I-f\,\left(\tens C_1-\tens C_0\right)\dbldot\langle\tens A_1(\vec n)\rangle\dbldot\tens P_d\right]^{-1}\dbldot\left(\tens C_1-\tens C_0\right)\dbldot\langle\tens A_1(\vec n)\rangle\dbldot\Eb
\eeqal
The effective stiffness tensor is given by
\beqal
\label{PCW_stiffness}
\tens C^{\mathrm{PCW}}=\tens C_0+f\,\left[\tens I-f\,\left(\tens C_1-\tens C_0\right)\dbldot\langle\tens A_1(\vec n)\rangle\dbldot\tens P_d\right]^{-1}\dbldot\left(\tens C_1-\tens C_0\right)\dbldot\langle\tens A_1(\vec n)\rangle
\eeqal
Here again, the average $\langle\tens A_1\rangle$ is necessary to compute this homogenized stiffness.

\section{Expression of $\langle\tens A_1\rangle$ using Walpole basis and orientation tensors}
We use a suited basis for fourth-order tensors with transverse isotropic symmetry.
This basis introduced by Walpole \cite{Walpole} is presented in Appendix \ref{Walpole_basis}, with the associated notations.
Indeed, the localisation tensor $\tens A_1$ of a spheroidal inclusion is transversely isotropic. We begin by the expression of Hill tensor
in Walpole basis, then we express the elastic tensors in this basis, and finally we give the expression of $\langle\tens A_1\rangle$ by using orientation
tensors.

\subsection{Hill tensor}
The transversely isotropic Hill tensor in Walpole basis is noted as
\beqal
\tens P(\vec n)=\left(\tens p,p_F,p_G\right)\quad\text{with}\quad\tens p=\begin{pmatrix}p_1&p_3\\p_4&p_2\end{pmatrix}
\eeqal
Following Parnell \cite{Parnell}, we write
\beqal
\gamma_e=\dfrac1{1-e^2}-\dfrac{e}{1-e^2}\dfrac{1}{\sqrt{|e^2-1|}}\mathrm{Argch}(e)\qquad\text{and}\quad\gamma=\dfrac12(1-\gamma_e)
\eeqal
where $e$ is the aspect ratio of the spheroid (which may be greater or lower than $1$), and (we adapted formulas (5.49)-(5.53) from \cite{Parnell}):
\beqal
&\psi_1=\dfrac{3\gamma-1}{2(1-e^2)}\qquad\text{and}\quad\psi_2=\dfrac{e^2(4\gamma-1)-\gamma}{4(1-e^2)}\\
&\psi_3=\dfrac{e^2(1-2\gamma)-\gamma}{4(1-e^2)}
\eeqal
The components of Hill tensor are then
\beqal
&p_1=\dfrac1{\mu_0}\left(1-2\gamma+\dfrac1{1-\nu_0}\psi_1\right)\quad p_2=\dfrac1{\mu_0}\left(\gamma+\dfrac1{1-\nu_0}\psi_2\right)\quad p_3=p_4=\dfrac{\sqrt2\,\psi_3}{\mu_0(1-\nu_0)}\\
&p_F=\dfrac1{\mu_0}\left(\gamma+\dfrac1{2(1-\nu_0)}\psi_2\right)\quad p_G=\dfrac1{\mu_0}\left(\dfrac{1-\gamma}2+\dfrac2{1-\nu_0}\psi_3\right)
\eeqal

These expressions allow to recast $\tens P(\vec n)$ into
\beqal
\label{Hill_geom}
\tens P(\vec n)=\dfrac1{\mu_0}\tens Q_e(\vec n)+\dfrac{6k_0+2\mu_0}{\mu_0(3k_0+4\mu_0)}\tens R_e(\vec n)\qquad\left(\text{note that}\quad\dfrac{6k_0+2\mu_0}{\mu_0(3k_0+4\mu_0)}=\dfrac1{\mu_0(1-\nu_0)}\right)
\eeqal

where $\tens Q_e(\vec n)$ and $\tens R_e(\vec n)$ are geometrical quantities given in Walpole basis by
\beqal
&\tens Q_e(\vec n)=\left(\tens q,q_F,q_G\right)\qquad\text{and}\quad\tens R_e(\vec n)=\left(\tens r,r_F,r_G\right)
\eeqal

with
\beqal
\label{Q_e_R_e}
&\tens q=\begin{pmatrix}1-2\gamma&0\\0&\gamma\end{pmatrix}\quad\tens r=\begin{pmatrix}\psi_1&\sqrt2\psi_3\\\sqrt2\psi_3&\psi_2\end{pmatrix}\\
&q_F=\gamma,\quad q_G=\dfrac{1-\gamma}2,\quad r_F=\psi_2/2,\quad r_G=2\psi_3
\eeqal

\subsection{Elastic tensors}
The elastic tensors can be written as
\beqal
\tens C_1=3k_1\tens J+2\mu_1\tens K\qquad\tens C_0=3k_0\tens J+2\mu_0\tens K
\eeqal
and from the expressions of $\tens J$ and $\tens K$ in Walpole basis we can write $\tens C_i$ ($i=0,1$) in this basis:
\beqal
\tens C_i=\left(\tens c_i,2\mu_i,2\mu_i\right)\quad\text{with}\quad\tens c_i=3k_i\,\tens j+2\mu_i\,\tens k
\eeqal
Hence we will denote $\tens C_1-\tens C_0$ by
\beqal
\tens \delta\tens C = \left(\tens\delta\tens c,2\delta\mu,2\delta\mu\right)\quad\text{with}\quad\tens\delta\tens c=3\delta k\,\tens j+2\delta\mu\,\tens k=\dfrac13\begin{pmatrix}3\,\delta k+4\,\delta\mu&\sqrt2\,(3\,\delta k-2\,\delta\mu)\\\sqrt2\,(3\,\delta k-2\,\delta\mu)&6\,\delta k+2\,\delta\mu\end{pmatrix}
\eeqal
where the $\delta$ is sufficiently explicit. Note that the basis used here can be associated to any direction $\vec n$,
so that in the following we will choose the most relevant direction.

\subsection{Expression of $\langle\tens A_1\rangle$}
We use the Walpole basis associated to the normal $\tens n$ of one given inclusion.
\beqal
\tens I+\tens P(\vec n)\dbldot\left(\tens C_1-\tens C_0\right)=\left(\tens 1+\tens p\dbldot\tens\delta\tens c,1+2\, p_F\delta\mu,1+2\,p_G\delta\mu\right)
\eeqal
Its inverse is
\beqal
\tens A_1(\tens n)=\left(\tens a,1/\left(1+2\,p_F\delta\mu\right),1/\left(1+2\,p_G\delta\mu\right)\right)\quad\text{with}\quad\tens a =\left(\tens i+\tens p\dbldot\tens\delta\tens c\right)^{-1}
\eeqal
Hence, the average $\langle\tens A_1\rangle$ is given by
\beqal
\label{A_average}
\langle\tens A_1\rangle=a_1 \langle\tens E_1\rangle + a_2 \langle\tens E_2\rangle + a_3 \langle\tens E_3\rangle + a_4 \langle\tens E_4\rangle + a_F \langle\tens F\rangle+a_G \langle\tens G\rangle
\eeqal
where
\beqal
\label{formulas_aFaG}
&\tens a =\begin{pmatrix}a_1&a_3\\a_4&a_2\end{pmatrix}\\
&a_F=1/\left(1+2\,p_F\delta\mu\right),\quad a_G =1/\left(1+2\,p_G\delta\mu\right)
\eeqal
The inverse of $\tens a$ is given by
\beqal
\label{a_inv}
\tens a^{-1}=\left(\tens i+\tens p\dbldot\tens\delta\tens c\right)=\begin{pmatrix}b_1&b_3\\b_4&b_2\end{pmatrix}
\eeqal
with
\beqal
\label{bi}
&b_1=1+\dfrac13\left(p_1\left(3\,\delta k+4\,\delta\mu\right)+p_3\sqrt2\,\left(3\,\delta k-2\,\delta\mu\right)\right)\\
&b_2=1+\dfrac13\left(p_4\sqrt2\,\left(3\,\delta k-2\,\delta\mu\right)+p_2\,\left(6\,\delta k+2\,\delta\mu\right)\right)\\
&b_3=\dfrac13\left(p_1\sqrt2\,\left(3\,\delta k-2\,\delta\mu\right)+p_3\,\left(6\,\delta k+2\,\delta\mu\right)\right)\\
&b_4=\dfrac13\left(p_4\left(3\,\delta k+4\,\delta\mu\right)+p_2\,\sqrt2\,\left(3\,\delta k-2\,\delta\mu\right)\right)\\
\eeqal
and hence
\beqal
\label{formulas_ai}
a_1=b_2/\Delta,\quad a_2=b_1/\Delta,\quad a_3=-b_3/\Delta,\quad a_4=-b_4/\Delta\quad\text{with}\quad\Delta=b_1\,b_2-b_3\,b_4
\eeqal
The average $\langle\tens A_1\rangle$ can hence be computed with \eqref{A_average} together with the averages of the Walpole basis tensors given by formulas \eqref{app_averages} and \eqref{app_averages_FG}.
These averages are computed by means of the orientation tensors, defined in Appendix \ref{sec_app_averages}.
Once obtained, the formulas \eqref{MT_stiffness} and \eqref{PCW_stiffness} give the corresponding homogenized stiffnesses according to Mori-Tanaka scheme and PCW scheme.
We now turn to the expressions of the derivatives of the homogenized stiffnesses w.r.t. the material parameters.

\section{Derivatives and second-moments}
The second-moments of the strains are the averages $\langle\tepsilon\otimes\tepsilon\rangle_r$ ($r=0,1$), which are helpful to compute the fluctuations of $\tepsilon$
on phase $r$, by means of the K$\oe$nig-Huygens formula:
\beqal
\langle\left(\tepsilon-\langle\tepsilon\rangle_r\right)\otimes\left(\tepsilon-\langle\tepsilon\rangle_r\right)\rangle_r=\langle\tepsilon\otimes\tepsilon\rangle_r-\langle\tepsilon\rangle_r\otimes\langle\tepsilon\rangle_r
\eeqal
This fluctuation is also useful in non-linear homogenization (see \cite{Nonlinearcomposites}). The second-moments can be expressed as the derivative of the homogenized stiffness (see \cite{kreher,suquet96}).
\beqal
\langle\tepsilon_{ij}\tepsilon_{kl}\rangle_r=\dfrac1{c_r}\Eb\dbldot\deriv{\tens C^{\hom}}{C^{ijkl}_r}\dbldot\Eb
\eeqal
Here we consider locally isotropic phases, so that every elastic tensor will be represented only by its two moduli $k,\mu$. Hence we obtain the following second-moments:
\beqal
\label{isotropic_moments}
&\langle\varepsilon_m^2\rangle_r=\dfrac1{9c_r}\Eb\dbldot\deriv{\tens C^{\hom}}{k_r}\dbldot\Eb\\
&\langle\tepsilon_{\mathrm{eq}}^2\rangle_r=\dfrac1{3c_r}\Eb\dbldot\deriv{\tens C^{\hom}}{\mu_r}\dbldot\Eb
\eeqal
where $\tepsilon_{\mathrm{eq}}^2=\dfrac23\tepsilon_d\dbldot\tepsilon_d$, with $\tepsilon_d=\tens K\dbldot\tepsilon$, and $\varepsilon_m=\dfrac13\mathrm{tr}\,\tepsilon$.

\subsection{Mori-Tanaka scheme}
Mori-Tanaka homogenized stiffness \eqref{MT_stiffness} can be expressed as
\beqal
\tens C^{\mathrm{MT}}=\tens C^{\mathrm{DS}}\dbldot\tens A_{\mathrm{MT}}^{-1}\quad\text{with}\quad\tens A_{\mathrm{MT}}=(1-f)\,\tens I + f\,\langle\tens A_1\rangle 
\eeqal
where $\tens C^{\mathrm{DS}}=(1-f)\,\tens C_0 + f\,\tens C_1\dbldot\langle\tens A_1\rangle$, is the homogenized stiffness given by the 'dilute scheme'.
The derivation w.r.t. any material parameter is then given by
\beqal
\label{dot_CMT}
\dot{\tens C}^{\mathrm{MT}}=\dot{\tens C}^{\mathrm{DS}}\dbldot\tens A_{\mathrm{MT}}^{-1}-f\,\tens C^{\mathrm{DS}}\dbldot\tens A_{\mathrm{MT}}^{-1}\dbldot\langle\dot{\tens A}_1\rangle \dbldot\tens A_{\mathrm{MT}}^{-1}
\eeqal
where the $\,\dot{}\,$ represents the derivative. The derivation of $\tens C^{\mathrm{DS}}$ is straightforward:
\beqal
\label{dot_CDS}
\dot{\tens C}^{\mathrm{DS}}=(1-f)\,\dot{\tens C}_0 + f\,\dot{\tens C}_1\dbldot\langle\tens A_1\rangle+ f\,\tens C_1\dbldot\langle\dot{\tens A}_1\rangle
\eeqal
We hence see that the derivative $\langle\dot{\tens A}_1\rangle$ is necessary to compute the derivative of $\tens C^{\mathrm{MT}}$.

\subsection{PCW scheme}
The formula \eqref{PCW_stiffness} of PCW stiffness can be recast into
\beqal
\label{C_PCW}
\tens C^{\mathrm{PCW}}=\tens C_0+f\,\tens A_{\mathrm{PCW}}^{-1}\dbldot\tens\delta\tens C\dbldot\langle\tens A_1\rangle\quad\text{with}\quad\tens\delta\tens C=\tens C_1-\tens C_0
\eeqal
with
\beqal
\label{A_PCW}
\tens A_{\mathrm{PCW}}=\tens I-f\,\tens\delta\tens C\dbldot\langle\tens A_1\rangle\dbldot\tens P_d
\eeqal
The derivation of \eqref{A_PCW} gives
\beqal
\label{dot_APCW}
\dot{\tens A}_{\mathrm{PCW}}=-f\,\tens\delta\dot{\tens C}\dbldot\langle\tens A_1\rangle\dbldot\tens P_d-f\,\tens\delta\tens C\dbldot\langle\dot{\tens A}_1\rangle\dbldot\tens P_d-f\,\tens\delta\tens C\dbldot\langle\tens A_1\rangle\dbldot\dot{\tens P}_d
\eeqal
Besides, the derivative of \eqref{C_PCW} gives
\beqal
\label{dot_CPCW}
\dot{\tens C}^{\mathrm{PCW}}=\dot{\tens C}_0-f\,\tens A_{\mathrm{PCW}}^{-1}\dbldot\dot{\tens A}_{\mathrm{PCW}}\dbldot\tens A_{\mathrm{PCW}}^{-1}\dbldot\tens\delta\tens C\dbldot\langle\tens A_1\rangle+f\,\tens A_{\mathrm{PCW}}^{-1}\dbldot\tens\delta\dot{\tens C}\dbldot\langle\tens A_1\rangle+f\,\tens A_{\mathrm{PCW}}^{-1}\dbldot\tens\delta\tens C\dbldot\langle\dot{\tens A}_1\rangle
\eeqal
We hence see that the derivatives $\dot{\tens P}_d$ and $\langle\dot{\tens A}_1\rangle$ are necessary to compute the derivative of $\tens C^{\mathrm{PCW}}$.

\subsection{Expressions for the derivatives $\langle\dot{\tens A}_1\rangle$ and $\dot{\tens P}_d$}
The derivation of expression \eqref{Hill_geom} directly gives the derivative of a Hill tensor relative to a spheroid:
\beqal
\label{dot_P}
\dot{\tens P}(\vec n)=\dot{\alpha}_0\tens Q_e(\vec n)+\dot{\beta}_0\tens R_e(\vec n)
\eeqal
where
\beqal
\dot{\alpha}_0=-\dfrac{\dot{\mu_0}}{\mu_0^2}\quad\text{and}\quad\dot{\beta}_0=\dfrac{\left(6\dot{k}_0+2\dot{\mu}_0\right)\left(3k_0\mu_0+4\mu_0^2\right)-\left(6k_0+2\mu_0\right)\left(3k_0\dot{\mu}_0+3\dot{k}_0\mu_0+8\mu_0\,\dot{\mu}_0\right)}{\left(3k_0\mu_0+4\mu_0^2\right)^2}
\eeqal
These formulas hence provide the derivative $\dot{\tens P}_d$ by replacing the vector $\tens n$ by the vector which defines the spheroidal distribution (note that the result is no more valid when $\tens P_d$ is defined by an ellipsoid with three different axes, which is not a spheroid).

\medskip

The derivative $\langle\dot{\tens A}_1\rangle$ is obtained by derivation of \eqref{A_average}:
\beqal
\label{dot_A}
\langle\dot{\tens A}_1\rangle=\dot{a}_1 \langle\tens E_1\rangle + \dot{a}_2 \langle\tens E_2\rangle + \dot{a}_3 \langle\tens E_3\rangle + \dot{a}_4 \langle\tens E_4\rangle + \dot{a}_F \langle\tens F\rangle+\dot{a}_G \langle\tens G\rangle
\eeqal
because the averages of the tensors of Walpole basis are geometrical quantities which do not depend on the material parameters.
We have also
\beqal
\dot{a}_F=-2\dfrac{p_F\,\delta\dot{\mu}+\dot{p}_F\,\delta\mu}{(1+2p_F\,\delta\mu)^2}\quad\text{with}\quad\dot{p}_F=\dot{\alpha}_0\,q_F+\dot{\beta}_0\,r_F\\
\dot{a}_G=-2\dfrac{p_G\,\delta\dot{\mu}+\dot{p}_G\,\delta\mu}{(1+2p_G\,\delta\mu)^2}\quad\text{with}\quad\dot{p}_G=\dot{\alpha}_0\,q_G+\dot{\beta}_0\,r_G
\eeqal
Derivation of \eqref{formulas_ai} gives
\beqal
&\dot{a}_1=(\dot{b}_2\,\Delta-b_2\,\dot{\Delta})/\Delta^2,\quad \dot{a}_2=(\dot{b}_1\,\Delta-b_1\,\dot{\Delta})/\Delta^2\\
&\dot{a}_3=-(\dot{b}_3\,\Delta-b_3\,\dot{\Delta})/\Delta^2,\quad \dot{a}_4=-(\dot{b}_4\,\Delta-b_4\,\dot{\Delta})/\Delta^2\\
&\dot{\Delta}=\dot{b}_1\,b_2+b_1\,\dot{b}_2-\dot{b}_3\,b_4-b_3\,\dot{b}_4
\eeqal
and derivation of \eqref{bi} gives
\beqal
&\dot{b}_1=\dfrac13\left(\dot{p}_1\left(3\,\delta k+4\,\delta\mu\right)+\dot{p}_3\sqrt2\,\left(3\,\delta k-2\,\delta\mu\right)+p_1\left(3\,\delta \dot k+4\,\delta\dot \mu\right)+p_3\sqrt2\,\left(3\,\delta \dot k-2\,\delta\dot \mu\right)\right)\\
&\dot{b}_2=\dfrac13\left(\dot{p}_4\sqrt2\,\left(3\,\delta k-2\,\delta\mu\right)+\dot{p}_2\,\left(6\,\delta k+2\,\delta\mu\right)+p_4\sqrt2\,\left(3\,\delta \dot k-2\,\delta\dot \mu\right)+p_2\,\left(6\,\delta \dot k+2\,\delta\dot \mu\right)\right)\\
&\dot{b}_3=\dfrac13\left(\dot{p}_1\sqrt2\,\left(3\,\delta k-2\,\delta\mu\right)+\dot{p}_3\,\left(6\,\delta k+2\,\delta\mu\right)+p_1\sqrt2\,\left(3\,\delta \dot k-2\,\delta\dot \mu\right)+p_3\,\left(6\,\delta \dot k+2\,\delta\dot \mu\right)\right)\\
&\dot{b}_4=\dfrac13\left(\dot{p}_4\left(3\,\delta k+4\,\delta\mu\right)+\dot{p}_2\,\sqrt2\,\left(3\,\delta k-2\,\delta\mu\right)+p_4\left(3\,\delta \dot k+4\,\delta\dot \mu\right)+p_2\,\sqrt2\,\left(3\,\delta \dot k-2\,\delta\dot \mu\right)\right)\\
\eeqal
whereas the derivatives $\dot{p}_i$ ($i=1,...,4$) are given by
\beqal
\dot{p}_i=\dot{\alpha}_0\,q_i+\dot{\beta}_0\,r_i
\eeqal
where $q_i,r_i$ ($i=1,...,4$) are the components of $\tens q,\tens r$ introduced by \eqref{Q_e_R_e}.

\medskip

The second-moments of the strains are hence given by \eqref{isotropic_moments}, and the derivatives of $\tens C^{\mathrm{hom}}$ are given by \eqref{dot_CMT} and \eqref{dot_CDS} for Mori-Tanaka
scheme, and \eqref{dot_APCW} and \eqref{dot_CPCW} for PCW scheme. Besides, these formulas also depend on the derivatives $\dot{\tens P}_d$ (provided by \eqref{dot_P}) and $\langle\dot{A}_1\rangle$ (provided by \eqref{dot_A}).
For each expression, the tuple $(\dot{k}_0,\dot{k}_1,\dot{\mu}_0,\dot{\mu}_1)$ must corresponds to the derivation performed: $(\dot{k}_0,\dot{k}_1,\dot{\mu}_0,\dot{\mu}_1)=(1,0,0,0)$ for a derivation
w.r.t. $k_0$, and so on.
We now turn to more specific distributions of orientations.

\section{Specific distributions}
\subsection{Isotropic distribution}
We assume here an isotropic distribution of orientations. Following \ref{app_iso}, we have
\beqal
\langle\tens A_1\rangle&=\dfrac13\left(a_1+2a_2+\sqrt2\left(a_3+a_4\right)\right)\tens J + \dfrac1{15}\left(2a_1+a_2-\sqrt2\left(a_3+a_4\right)+6a_F+6a_G\right)\tens K\\
&\equiv a_J\tens J+a_K\tens K
\eeqal
This leads to the following expression of Mori-Tanaka homogenized isotropic moduli:
\beqal
\label{iso_MT}
k^{\mathrm{MT}}=\dfrac{k_0+f\,(k_1a_J-k_0)}{1+f\,(a_J-1)}\\
\mu^{\mathrm{MT}}=\dfrac{\mu_0+f\,(\mu_1a_K-\mu_0)}{1+f\,(a_K-1)}
\eeqal
Now, for Ponte-Casta\~neda and Willis scheme, we take $\tens P_d$ equal to the Hill tensor of a sphere:
\beqal
\tens P_d=\dfrac1{3k_0+4\mu_0}\tens J+\dfrac{3k_0+6\mu_0}{5\mu_0(3k_0+4\mu_0)}\tens K\equiv p_J\,\tens J+p_K\,\tens K
\eeqal
so that
\beqal
\label{iso_PCW}
k^{\mathrm{PCW}}=k_0+f\,\dfrac{a_J\,(k_1-k_0)}{1-3f\,a_J\,p_J\,(k_1-k_0)}\\
\mu^{\mathrm{PCW}}=\mu_0+f\,\dfrac{a_K\,(\mu_1-\mu_0)}{1-2f\,a_K\,p_K\,(\mu_1-\mu_0)}
\eeqal

\subsection{Planar isotropic distribution}
We now assume a planar isotropic distribution of orientations. Following \ref{app_Tiso}, we can express the average $\langle\tens A_1\rangle$ in
the Walpole basis associated to \emph{the distribution}.
\beqal
\langle\tens A_1\rangle&=\left(\tens \alpha,\alpha_F,\alpha_G\right)\quad\text{with}\quad\tens \alpha=\begin{pmatrix}\alpha_1&\alpha_3\\\alpha_4&\alpha_2\end{pmatrix}\\
\eeqal
with
\beqal
&\alpha_1=\dfrac12a_2+a_F,\quad\alpha_2=\dfrac12a_1+\dfrac14a_2+\dfrac{\sqrt2}4\left(a_3+a_4\right)-\dfrac12a_F\\
&\alpha_3=\dfrac{\sqrt2}4a_2+\dfrac12a_4,\quad\alpha_4=\dfrac{\sqrt2}4a_2+\dfrac12a_3\\
&\alpha_F=\dfrac14\left(a_1+a_2\right)-\dfrac{\sqrt2}8\left(a_3+a_4\right)+\dfrac14a_F+\dfrac12a_G,\quad\alpha_G=\dfrac12\left(a_F+a_G\right)
\eeqal

\subsubsection{Mori-Tanaka scheme for planar isotropic distribution}
\label{MT_planar}
We hence have, in Walpole basis relative to the transverse isotropic distribution,
\beqal
&\tens C_0 + f\,\left(\tens C_1\dbldot\langle\tens A_1\rangle-\tens C_0\right)=\left(\tens c_0+f\,(\tens c_1\dbldot\tens \alpha-\tens c_0),2\mu_0+f\,(2\mu_1\alpha_F-2\mu_0),2\mu_0+f\,(2\mu_1\alpha_G-2\mu_0)\right)\\
&\tens I+f\,\left(\langle\tens A_1\rangle-\tens I\right)=\left(\tens i+f\,(\tens \alpha-\tens i),1+f\,(\alpha_F-1),1+f\,(\alpha_G-1)\right)
\eeqal
so that Mori-Tanaka effective stiffness is
\beqal
&\tens C^{\mathrm{MT}}=\left(\tens c^{\mathrm{MT}},2\mu^{\mathrm{MT}}_{T},2\mu^{\mathrm{MT}}_{P}\right)
\eeqal
with (note the similarity with \eqref{iso_MT})
\beqal
&\tens c^{\mathrm{MT}}=\left[\tens c_0+f\,(\tens c_1\dbldot\tens \alpha-\tens c_0)\right]\dbldot\left[\tens i+f\,(\tens \alpha-\tens i)\right]^{-1}\\
&\mu^{\mathrm{MT}}_{T}=\dfrac{\mu_0+f\,(\mu_1\alpha_F-\mu_0)}{1+f\,(\alpha_F-1)},\quad\text{and}\quad\mu^{\mathrm{MT}}_{P}=\dfrac{\mu_0+f\,(\mu_1\alpha_G-\mu_0)}{1+f\,(\alpha_G-1)}
\eeqal

\subsubsection{PCW scheme for planar isotropic distribution}
\label{PCW_planar}
$\tens P_d$ is the Hill tensor associated to a spheroid, defined by the direction $\tens n^d$.
Hence, we can work in the associated Walpole basis and write:
\beqal
\tens P_d=\left(\tens p^d,p_F^d,p_G^d\right)
\eeqal
This gives the following expression of the effective PCW stiffness:
\beqal
&\tens C^{\mathrm{PCW}}=\left(\tens c^{\mathrm{PCW}},2\mu^{\mathrm{PCW}}_{T},2\mu^{\mathrm{PCW}}_{P}\right)
\eeqal
with (note the similarity with \eqref{iso_PCW})
\beqal
&\tens c^{\mathrm{PCW}}=\tens c_0+f\,\left[\tens i-f\,(\tens c_1-\tens c_0)\dbldot\tens \alpha\dbldot\tens p^d\right]^{-1}\dbldot(\tens c_1-\tens c_0)\dbldot\tens \alpha\\
&\mu^{\mathrm{PCW}}_{T}=\mu_0+f\dfrac{\alpha_F\,(\mu_1-\mu_0)}{1-2f\,\alpha_F\,p_F^d\,(\mu_1-\mu_0)},\quad\text{and}\quad\mu^{\mathrm{PCW}}_{P}=\mu_0+f\dfrac{\alpha_G\,(\mu_1-\mu_0)}{1-2f\,\alpha_G\,p_G^d\,(\mu_1-\mu_0)}
\eeqal

\subsection{Unidirectional distribution}
In this case, $\langle\tens A_1\rangle$ is just equal to $\tens A_1$ so that
\beqal
\langle\tens A_1\rangle=\left(\tens a,a_F,a_G\right)
\eeqal
in the Walpole basis relative to the unique direction.
Formulas of Secs.~\ref{MT_planar} and \ref{PCW_planar} can be conserved by replacing $\tens \alpha$ by $\tens a$, and $\alpha_F,\alpha_G$ by $a_F,a_G$.

\appendix
\section{Walpole basis}
\label{Walpole_basis}
Walpole \cite{Walpole} used different bases correponding to a wide range of symmetries of the fourth-order tensors.
Here, we are interested on the basis for transverse isotropic cases where the privileged direction is noted $\tens n$.
We first introduce the second-order projectors
\beqal
\tens p = \tens n\otimes \tens n\qquad \tens q = \tens 1 -  \tens n\otimes \tens n
\eeqal 
where $\tens 1$ is the second-order identity tensor. Then, the basis of transverse isotropic fourth-order tensors is formed by six tensors:
\beqal
&\tens E_1(\tens n)=\tens p\otimes \tens p\qquad\tens E_2(\tens n)=\dfrac12\tens q\otimes \tens q\\
&\tens E_3(\tens n)=\dfrac{1}{\sqrt2}\tens p\otimes \tens q\qquad\tens E_4(\tens n)=\dfrac{1}{\sqrt2}\tens q\otimes \tens p\\
&\tens F(\tens n)=\tens q\,\underline{\overline{\otimes}}\,\tens q-\dfrac12\tens q\otimes \tens q\qquad\tens G(\tens n)=\tens p\,\underline{\overline{\otimes}}\,\tens q+\tens q\,\underline{\overline{\otimes}}\,\tens p\\
\eeqal
where 
\beqal
\left(\tens a\,\underline{\overline{\otimes}}\,\tens b\right)_{ijkl} = \dfrac12 \left(a_{ik}b_{jl}+a_{il}b_{jk} \right)
\eeqal
With this kind basis, the following multiplication table is provided by Walpole:

\medskip

\begin{center}
\begin{tabular}{ c | c | c | c | c | c | c }
           &$\tens E_1$&$\tens E_2$&$\tens E_3$&$\tens E_4$&$\tens F$&$\tens G$\\
           \hline
$\tens E_1$&$\tens E_1$&$\tens 0$&$\tens E_3$&$\tens 0$&$\tens 0$&$\tens 0$\\
$\tens E_2$&$\tens 0$&$\tens E_2$&$\tens 0$&$\tens E_4$&$\tens 0$&$\tens 0$\\
$\tens E_3$&$\tens 0$&$\tens E_3$&$\tens 0$&$\tens E_1$&$\tens 0$&$\tens 0$\\
$\tens E_4$&$\tens E_4$&$\tens 0$&$\tens E_2$&$\tens 0$&$\tens 0$&$\tens 0$\\
$\tens F$&  $\tens 0$&$\tens 0$&$\tens 0$&$\tens 0$&$\tens F$&$\tens 0$\\
$\tens G$&  $\tens 0$&$\tens 0$&$\tens 0$&$\tens 0$&$\tens 0$&$\tens G$
\end{tabular}
\end{center}
A transverse isotropic tensor in Walpole basis
\beqal
&\tens B = \beta_1\tens E_1(\vec n)+\beta_2\tens E_2(\vec n)+\beta_3\tens E_3(\vec n)+\beta_4\tens E_4(\vec n)+\beta_F\tens F(\vec n)+\beta_G\tens G(\vec n)\\
\eeqal
can be represented by:
\beqal
\tens B = \left(\tens \beta,\beta_F,\beta_G\right)\quad\text{with}\quad\tens \beta=\begin{pmatrix}\beta_1&\beta_3\\\beta_4&\beta_2\end{pmatrix}
\eeqal
so that the multiplication of two tensors $\tens B$ and $\tens B'$ is obtained by multiplication of each component (the multiplication of the first component being a matrix product)
and the inverse of $\tens B$ is given by
\beqal
&\tens B^{-1}=\left(\tens \beta^{-1},1/\beta_F,1/\beta_G\right)
\eeqal

Note also that we have the following decomposition of tensors $\tens I,\tens J$ and $\tens K$ in this particular basis:
\beqal
&\tens J=\left(\tens j,0,0\right)\quad\text{with}\quad\tens j=\dfrac13\begin{pmatrix}1&\sqrt2\\\sqrt2&2\end{pmatrix}\quad\text{and}\quad\tens K=\left(\tens k,1,1\right)\quad\text{with}\quad\tens k=\dfrac13\begin{pmatrix}2&-\sqrt2\\-\sqrt2&1\end{pmatrix}\\
&\tens I=\left(\tens i,1,1\right)\quad\text{with}\quad\tens i=\begin{pmatrix}1&0\\0&1\end{pmatrix}
\eeqal
Note that the notation $\mathcal H^1,...,\mathcal H^6$ used by Parnell \cite{Parnell} is obtained by
doing the following changes:
\beqal
&\mathcal H^4\leftrightarrow\tens E_1\quad\mathcal H^1\leftrightarrow\tens E_2\quad\dfrac1{\sqrt2}\mathcal H^3\leftrightarrow\tens E_3\\
&\dfrac1{\sqrt2}\mathcal H^2\leftrightarrow\tens E_4\quad\mathcal H^5\leftrightarrow\tens F\quad\mathcal H^6\leftrightarrow\tens G
\eeqal

\section{Averages of Walpole basis tensors using orientation tensors}
\label{sec_app_averages}
We note the following orientation tensors:
\beqal
&\tens A_2=\langle\vec n\otimes\vec n\rangle\\
&\tens A_4=\langle\vec n\otimes\vec n\otimes\vec n\otimes\vec n\rangle
\eeqal
The averages are taken over all inclusions. More particularly, let us define a normal
$\vec n(\theta,\varphi) = (\sin\theta\cos\varphi,\sin\theta\sin\varphi,\cos\theta)$ with the two usual angles $\theta,\varphi$.
We note the probability to find, in our distribution, a normal between $(\theta,\varphi)$ and $(\theta+\mathrm{d}\theta,\varphi+\mathrm{d}\varphi)$:
\beqal
\rho\,(\theta,\varphi)\,\mathrm{d}\theta\,\mathrm{d}\varphi
\eeqal
where $\rho$ is called the density probability. The orientation tensors are then given by
\beqal
&\tens A_2=\dfrac{1}{2\pi}\int_{\substack{\theta\in[0,\frac{\pi}2]\\\varphi\in[0,2\pi]}}\vec n(\theta,\varphi)\otimes\vec n(\theta,\varphi)\,\rho\,(\theta,\varphi)\,\mathrm{d}\theta\,\mathrm{d}\varphi\\
&\tens A_4=\dfrac{1}{2\pi}\int_{\substack{\theta\in[0,\frac{\pi}2]\\\varphi\in[0,2\pi]}}\vec n(\theta,\varphi)\otimes\vec n(\theta,\varphi)\otimes\vec n(\theta,\varphi)\otimes\vec n(\theta,\varphi)\,\rho\,(\theta,\varphi)\,\mathrm{d}\theta\,\mathrm{d}\varphi
\eeqal

These tensors permit to write the averages of each tensor of the Walpole basis related to $\vec n$. This gives
\beqal
\label{app_averages}
&\langle\tens E_1\rangle=\tens A_4\\
&\langle\tens E_2\rangle=\dfrac12\left(3\tens J-\tens A_2\otimes\tens 1-\tens 1\otimes\tens A_2+\tens A_4\right)\\
&\langle\tens E_3\rangle=\dfrac1{\sqrt2}\left(\tens A_2\otimes\tens 1-\tens A_4\right)\\
&\langle\tens E_4\rangle=\dfrac1{\sqrt2}\left(\tens 1\otimes\tens A_2-\tens A_4\right)\\
&\langle\tens F\rangle=\tens I-\tens 1\overline{\underline{\otimes}}\tens A_2-\tens A_2\overline{\underline{\otimes}}\tens 1 +\langle\tens n\otimes\tens n\overline{\underline{\otimes}}\tens n\otimes\tens n\rangle-\dfrac12\left(3\tens J-\tens A_2\otimes\tens 1-\tens 1\otimes\tens A_2+\tens A_4\right)\\
&\langle\tens G\rangle=\tens A_2\overline{\underline{\otimes}}\tens 1+\tens 1\overline{\underline{\otimes}}\tens A_2-2\langle\tens n\otimes\tens n\overline{\underline{\otimes}}\tens n\otimes\tens n\rangle
\eeqal
and given that $\tens n\otimes\tens n\overline{\underline{\otimes}}\tens n\otimes\tens n=\tens n\otimes\tens n\otimes\tens n\otimes \tens n$, we have
\beqal
\label{app_averages_FG}
&\langle\tens F\rangle=\tens I-\tens 1\overline{\underline{\otimes}}\tens A_2-\tens A_2\overline{\underline{\otimes}}\tens 1 -\dfrac12\left(3\tens J-\tens A_2\otimes\tens 1-\tens 1\otimes\tens A_2-\tens A_4\right)\\
&\langle\tens G\rangle=\tens A_2\overline{\underline{\otimes}}\tens 1+\tens 1\overline{\underline{\otimes}}\tens A_2-2\tens A_4
\eeqal

\subsection{Isotropic distribution}
\label{app_iso}
We assume the following density probability:
\beqal
\rho\,(\theta,\varphi)=\sin\theta
\eeqal
so that
\beqal
\tens A_2=\dfrac13\tens 1\quad\text{and}\quad\tens A_4=\dfrac13\tens J +\dfrac2{15}\tens K
\eeqal
In this case, we have
\beqal
&\langle\tens E_1\rangle=\dfrac13\tens J +\dfrac2{15}\tens K\\
&\langle\tens E_2\rangle=\dfrac23\tens J +\dfrac1{15}\tens K\\
&\langle\tens E_3\rangle=\dfrac{\sqrt2}3\tens J -\dfrac{\sqrt2}{15}\tens K\\
&\langle\tens E_4\rangle=\dfrac{\sqrt2}3\tens J -\dfrac{\sqrt2}{15}\tens K\\
&\langle\tens F\rangle=\dfrac25\tens K\\
&\langle\tens G\rangle=\dfrac25\tens K
\eeqal

\subsection{Planar isotropic distribution}
\label{app_Tiso}
We first define the direction $\tens n^d$ associated to the direction of transverse isotropy (indeed, a planar isotropic
distribution is transverse isotropic and the related axis is the vector normal to the plane to which belong
the axes of the inclusions).
We assume the following density probability:
\beqal
\rho\,(\theta,\varphi)=\delta_{\theta=\theta_0+\frac{\pi}2}
\eeqal
where $\delta_{\theta=\theta_0+\frac{\pi}2}$ is a Dirac associated to $\theta =\theta_0+\frac{\pi}2$, where $\theta_0$ is associated to $\vec n_d=(\sin\theta_0\cos\varphi_0,\sin\theta_0\sin\varphi_0,\cos\theta_0)$. This means that the integration can be performed only on $\varphi$.
The orientation tensors can be expressed in the Walpole basis relative to the vector $\tens n^d$:
\beqal
\tens A_2=\dfrac12\left(\tens 1-\tens n_d\otimes\tens n_d\right)\equiv\dfrac12\tens q^d
\eeqal
and
\beqal
\tens A_4=\dfrac12\tens E_2^d+\dfrac14\tens F^d
\eeqal
In this case, we have
\beqal
&\langle\tens E_1\rangle=\dfrac12\tens E_2^d+\dfrac14\tens F^d\\
&\langle\tens E_2\rangle=\dfrac12\tens E_1^d+\dfrac14\tens E_2^d+\dfrac{\sqrt2}4\left(\tens E_3^d+\tens E_4^d\right)+\dfrac14\tens F^d\\
&\langle\tens E_3\rangle=\dfrac{\sqrt2}4\tens E_2^d+\dfrac12\tens E_4^d-\dfrac{\sqrt2}8\tens F^d\\
&\langle\tens E_4\rangle=\dfrac{\sqrt2}4\tens E_2^d+\dfrac12\tens E_3^d-\dfrac{\sqrt2}8\tens F^d\\
&\langle\tens F\rangle=\tens E_1^d-\dfrac12\tens E_2^d+\dfrac14\tens F^d+\dfrac12\tens G^d\\
&\langle\tens G\rangle=\dfrac12\left(\tens F^d+\tens G^d\right)
\eeqal

\subsection{Unidirectional distribution}
We assume that $\vec n$ is fixed so that in this case, we simply have
\beqal
\tens A_2=\tens n\otimes\tens n=\tens p
\eeqal
and
\beqal
\tens A_4=\tens n\otimes\tens n\otimes\tens n\otimes\tens n=\tens E_1
\eeqal
In this case we simply have
\beqal
&\langle\tens E_1\rangle=\tens E_1,\quad\langle\tens E_2\rangle=\tens E_2,\quad\langle\tens E_3\rangle=\tens E_3\\
&\langle\tens E_4\rangle=\tens E_4,\quad\langle\tens F\rangle=\tens F,\quad\langle\tens G\rangle=\tens G
\eeqal

\end{document}